\documentclass[sigconf]{acmart}

\AtBeginDocument{%
  \providecommand\BibTeX{{%
    \normalfont B\kern-0.5em{\scshape i\kern-0.25em b}\kern-0.8em\TeX}}}

\usepackage{mdframed}
\usepackage{graphicx}
\usepackage{Templates/template}
\usepackage{multirow}
\usepackage{graphicx}
\usepackage{enumitem}
\usepackage{wrapfig}
\usepackage[capitalise]{cleveref}
\usepackage{array,multirow,graphicx}
\usepackage{hyperref}

\usepackage{xcolor}
\def\BibTeX{{\rm B\kern-.05em{\sc i\kern-.025em b}\kern-.08em
    T\kern-.1667em\lower.7ex\hbox{E}\kern-.125emX}}

\usepackage{amsmath}
\usepackage{booktabs} 
\usepackage{subcaption}
\usepackage{colortbl}
\usepackage{color}
\usepackage{listings}
\usepackage{balance}
\usepackage{colortbl}
\usepackage{tcolorbox}
\usepackage{url}
\usepackage{soul}
\usepackage{tcolorbox}
\usepackage{MnSymbol}
\usepackage{titlesec}
\titlespacing{\subsubsection}{0pt}{\parskip}{-\parskip}

\usepackage[font=small,labelfont=bf]{caption}
\usepackage{tcolorbox}
\definecolor{light-green}{rgb}{.5,1,.5}
\definecolor{light-pink}{rgb}{1,0.5,.5}

\floatstyle{ruled}
\newfloat{example}{thp}{lop}
\floatname{example}{Example }

\usepackage{xcolor}
\usepackage{listings}

\lstdefinestyle{yaml}{
     basicstyle=\color{blue}\footnotesize,
     rulecolor=\color{black},
     string=[s]{'}{'},
     stringstyle=\color{blue},
     comment=[l]{:},
     commentstyle=\color{black},
     morecomment=[l]{-},
     morecomment=[l]{+},
 }
 \newcommand{\RedL}{\makebox[0pt][l]{\color{red!6}\rule[-3pt]{\linewidth}{10pt}}}
\newcommand{\GreenL}{\makebox[0pt][l]{\color{green!6}\rule[-3pt]{\linewidth}{10pt}}}

\usepackage{listings}

\definecolor{codegreen}{rgb}{0,0.6,0}
\definecolor{codegray}{rgb}{0.5,0.5,0.5}
\definecolor{codepurple}{rgb}{0.58,0,0.82}
\definecolor{backcolour}{rgb}{0.95,0.95,0.92}

\lstdefinestyle{mystyle}{
  basicstyle=\ttfamily\footnotesize,
  breakatwhitespace=false, 
  breaklines=true,                 
  captionpos=b,                    
  xleftmargin=.01\textwidth, xrightmargin=.01\textwidth,
  numbers=left,                    
  numbersep=6pt,                  
  showspaces=false,                
  showstringspaces=false,
  showtabs=false,                  
  tabsize=1,
  frame=single,
}

\begin{document}
\renewcommand {\scriptsize} {\footnotesize}

\title{Empirical Analysis on CI/CD Pipeline Evolution in Machine Learning Projects}

\author{Dhia Elhaq Rzig}
\authornote{Equal Contribution}
\affiliation{
  \institution{University of Michigan - Dearborn}
  \country{Dearborn, MI, USA}
}
\email{dhiarzig@umich.edu}

\author{Alaa Houerbi}
\authornotemark[1] 
\affiliation{
  \institution{University of Michigan - Dearborn}
  \country{Dearborn, MI, USA}
}
\email{houerbi@umich.edu}

\author{
  Rahul Ghanshyam Chavan}
\affiliation{
  \institution{University of Michigan - Dearborn}
  \country{Dearborn, MI, USA}
}
\email{rchavan@umich.edu}



\author{Foyzul Hassan}
\affiliation{%
  \institution{University of Michigan - Dearborn}
  \country{Dearborn, MI, USA}
}
\email{foyzul@umich.edu}







\begin{abstract}

  The growing popularity of machine learning (ML) and the integration of ML components with other software artifacts has led to the use of continuous integration and delivery (CI/CD) tools, such as Travis CI, GitHub Actions, etc. that enable faster integration and testing for ML projects. Such CI/CD configurations and services require synchronization during the life cycle of the projects. 
  Several works discussed how CI/CD configuration and services change during their usage in traditional software systems. However, there is very limited knowledge of how CI/CD configuration and services change in ML projects.
  
To fill this knowledge gap, this work presents the first empirical analysis of how CI/CD configuration evolves for ML software systems. We manually analyzed 343 commits collected from 508 open-source ML projects to identify common CI/CD configuration change categories in ML projects and devised a taxonomy of 14 co-changes in CI/CD and ML components. Moreover, we developed a CI/CD configuration change clustering tool that identified frequent CI/CD configuration change patterns in 15,634 commits. Furthermore, we measured the expertise of ML developers who modify CI/CD configurations. Based on this analysis, we found that 61.8\% of commits include a change to the build policy and minimal changes related to performance and maintainability compared to general open-source projects. Additionally, the co-evolution analysis identified that CI/CD configurations, in many cases, changed unnecessarily due to bad practices such as the direct inclusion of dependencies and a lack of usage of standardized testing frameworks. More practices were found through the change patterns analysis consisting of using deprecated settings and reliance on a generic build language. Finally, our developer's expertise analysis suggests that experienced developers are more inclined to modify CI/CD configurations.
\end{abstract}


\maketitle
\bibliographystyle{IEEEtran}


\newcommand{\totalCountTestClasses}{523}
\newcommand{\totalCountTestMethods}{3051}
\newcommand{\totalCountTestLOCs}{68939}
\section{Introduction}
\vspace{-0.1cm}
Continuous Integration (CI)~\cite{beck2000extreme} establishes an automated way to build, package, and test software applications and encourages developers to commit code changes more frequently
~\cite{duvall2007continuous,hilton2016usage,beller2017oops}. 
After CI comes Continuous Delivery (CD), which automates the delivery of applications to selected environments in short cycles~\cite{humble2010continuous}.
CI/CD pipelines help create an automated and consistent process that helps reduce human errors, increase productivity in teams, and accelerate release cycles~\cite{duvall2007continuous,hilton2016usage, vasilescu2015quality}.
CI/CD has become the industry standard of modern software development ~\cite{shahin2017continuous} and has been widely adopted in Open-Source Software (OSS) projects ~\cite{hilton2016usage,staahl2014modeling} and in Machine Learning (ML) projects ~\cite{rzig2022characterizing} which have gained widespread popularity and significance in recent years ~\cite{gonzalez2020state,quantilus2020}. 
\\
Machine Learning shares common ground with traditional software development in the need for multiple iterations to enhance the quality of ML models. However, ML projects introduce a unique set of challenges due to their inherent complexity ~\cite{amershi2019software,haakman2021ai}. For instance, regular testing methods in CI can cause overfitting rendering accuracy measures unreliable for evaluating models  ~\cite{karlavs2020building}, and ML projects encounter challenges in version control and dependency management, leading to manual interventions during model experiments and deployments to address these issues ~\cite{lwakatare2020devops}. Furthermore, limited knowledge exists on software maintenance and evolution in the context of ML systems ~\cite{martinez2022software}.
A recent study by Zampetti et al. ~\cite{zampetti2021ci} explored how CI/CD configurations evolve over time in open-source projects and focused mostly on the restructuring actions occurring within these files. Other studies~\cite{Golzadeh20222,Kinsman2021HowDS,Widder2018ImLY} on CI/CD focus on the uses and challenges of utilizing CI/CD in traditional software systems. To the best of our knowledge, understanding the nuances of this co-evolution between ML and CI/CD changes remains an undiscovered territory. This research gap makes it challenging to discern the necessary adjustments in CI/CD configurations to effectively accommodate the changes in ML systems. 
\\
Building upon this context, this paper presents an in-depth analysis of the evolution of CI/CD configurations in ML projects. By examining the intricate relationships between changes in ML source files and corresponding adjustments in CI/CD configurations, analyzing the patterns of change, and evaluating developers' expertise, this study aims to unravel the complexities of sustaining and maintaining CI/CD setups for evolving ML models. 
We collected 508 open-source Python-based ML-enabled projects on GitHub having Travis CI as their CI/CD infrastructure. We opted for these filtering criteria since Python is the most popular language for ML-enabled projects, and Travis is their most popular CI tool~\cite{rzig2022characterizing}. We extract 15,634 commits from these projects modifying the ~\textit{.travis.yml} file. We filtered those commits to include only ones modifying both ML source files and CI/CD configurations and then we applied random sampling to obtain 343 commits which will be used for manual analysis. Through this work, we answer the following research questions:

\begin{itemize}[leftmargin=.1in,nolistsep,topsep=0in]
    \item \textbf{RQ1}: \textit{How do CI/CD pipelines evolve in ML projects?}\
    We observe that changes to the Build Policy in ML CI/CD configurations occur in over half of the commits, mostly due to updating the installation policy. Unlike Zampetti et al.’s ~\cite{zampetti2021ci} results for general-purpose projects, we found that Performance, Maintainability, and Build Process Organization are not major concerns when it comes to updating CI/CD configurations which may lead to technical debt and prolonged builds.

    \item \textbf{RQ2}: \textit{How do CI/CD pipelines co-evolve with ML code?}\
    We devise a taxonomy of 14 co-evolution categories and we find Testing and Dependency Management as the most prominent categories of change whereas Deployment and Data Versioning are infrequent. We identified two bad practices performed by ML developers like adding dependencies directly to the \textit{.travis.yml} file and not using the automatic discovery feature of testing frameworks.
    
    \item \textbf{RQ3}: \textit{What are the common patterns of change occurring in CI/CD configurations?}\
    We generated a comprehensive list of change patterns occurring in CI/CD pipelines in ML projects. The AST analysis supports our earlier findings and shows that there are minor adjustments related to deployment and failure handling. This is worrisome because it means that ML developers often resort to manual interventions for debugging build failure and for deploying models. 
    We found two more bad practices which consist of using deprecated Travis CI settings and relying on a generic build language.

    \item \textbf{RQ4}: \textit{How skilled are the developers changing CI/CD configurations in ML projects?} \
       Our analysis revealed a robust and statistically significant positive association between developers' project knowledge and expertise, and their involvement in modifying CI/CD configurations.
       This indicated that the more active and knowledgeable developers are more inclined to modify CI/CD configurations.
\end{itemize}

\noindent In summary, our study makes the following contributions:
\begin{itemize}[leftmargin=.1in,nolistsep,topsep=0in]
    \item The first quantitative and qualitative study of CI/CD configuration evolution in open-source ML projects.
    \item A taxonomy of 14 categories of co-changes between CI/CD configurations and ML source code.
    \item A list of common change patterns in CI/CD configurations in ML projects that can be used to mitigate challenges associated with ML CI evolution.
    \item A study on the expertise employed to change the pipeline configuration files. 
\end{itemize} 

Furthermore, our code scripts and empirical dataset are publicly available for researchers to replicate and build upon ~\cite{code}.

The remainder of this paper is organized as follows: ~\autoref{sec:background} discusses the background of the project, and ~\autoref{sec:methodology} outlines the research methodology employed in our empirical study. \autoref{sec:evaluation} provides a detailed analysis addressing the four research questions. Potential threats to the validity of our study are mentioned in ~\autoref{sec:discussion}. The research implications are discussed in ~\autoref{sec:implications}. Finally, ~\autoref{sec:related} delves into related works to our study, while ~\autoref{sec:conclusion} concludes this study.

\vspace{-0.2cm}
\section{Background}
\label{sec:background}

CI/CD pipelines have been used by different types of projects and have been adopted by a fair amount of ML projects ~\cite{rzig2022characterizing,Rzig_Hassan_Kessentini_2022}. In recent years, some CI/CD tools specifically designed for ML projects have emerged, including KubeFlow ~\cite{kubeflow}, Amazon Sagemaker ~\cite{sagemaker}, and Azure Machine Learning~\cite{azureml}. However, these ML-specific tools can not be used as standalone and typically complement traditional CI/CD tools for managing the codebase of ML-enabled projects. Despite the differences between traditional software development and ML projects, some CI/CD tools designed for the former, like Travis CI ~\cite{travisci}, remain popular in both domains. 
\\
Travis CI ~\cite{travisci} is one of the largest and most popular CI/CD services ~\cite{zampetti2021ci}. It supports a variety of programming languages and provides a cloud-based infrastructure, relieving developers of the burden of maintaining their own environments. 
Travis CI automatically detects changes to the repository on version control systems like GitHub and triggers a build based on predefined events. The build process is configured by the \textit{.travis.yml} file which resides in the root of the project repository. An example of a typical \textit{.travis.yml} file is shown in ~\autoref{listing:travis}.

\begin{lstlisting}[style=yaml, caption={An example of.travis.yml file.}, label={listing:travis},numbers=left,frame=lines,escapechar=\!]
language: python
os: 
  - linux
python:
  - 3.8
install:
  - pip install -r requirements.txt
script:
  - nosetests . --with-coverage
after_success:
  - codecov
\end{lstlisting}

\begin{figure}[!htbp]
    \centering
   \includegraphics[width=0.9\linewidth]{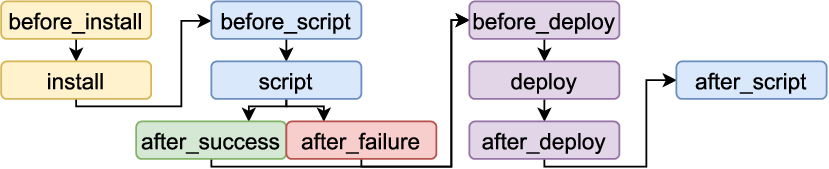}
   \vspace{-0.3cm}
    \caption{Travis CI job Lifecycle.}
    \vspace{-0.3cm}
    \label{fig:travis_jobs}
\end{figure}

The \textit{.travis.yml} file defines the build process through a series of stages. These stages consist of jobs running in parallel and executing different phases in a virtual environment. 
The build environment can be configured through the \texttt{os} keyword, which sets the Operating System of a job’s container, and the \texttt{language} keyword, which installs the tools and dependencies of a specific programming language. This configuration can be used multiple times and with different values to configure various environments for jobs, each executing the same sub-script in its designated environment.
The job executes a series of phases which are shown in ~\autoref{fig:travis_jobs}.

The typical phases include an \texttt{install} phase for dependency installation, a \texttt{script} phase for running tests, an \texttt{after\_success} phase to handle post-test actions such as coverage reporting, and a \texttt{deploy} phase for project deployment. In ~\autoref{listing:travis}, the configuration installs project dependencies using \texttt{pip}, runs tests with \texttt{nosetests}, and reports coverage to \texttt{codecov}.

\vspace{-0.2cm}
\section{Research Methodology}
\label{sec:methodology}

\begin{figure}[!htbp]
    \centering
   \includegraphics[width=\linewidth]{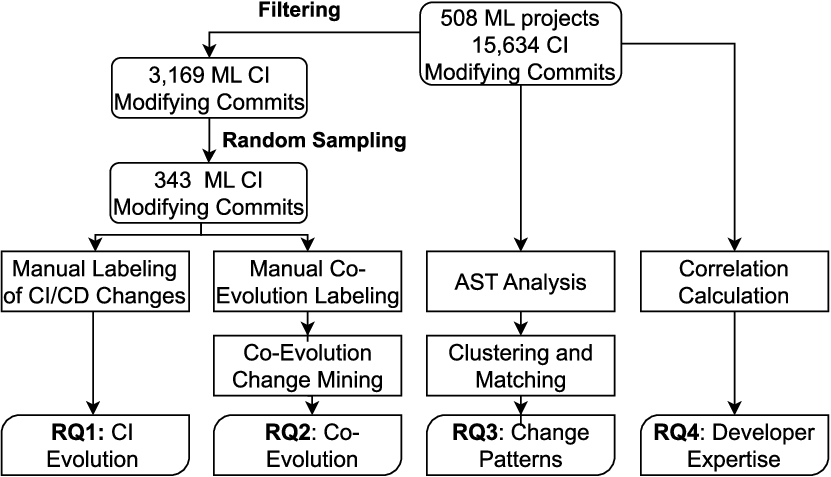}
   \vspace{-0.3cm}
    \caption{Overview of Research Approach}
    \vspace{-0.4cm}
    \label{fig:overall}
\end{figure}

~\autoref{fig:overall} shows an overview of the research methodology.
In this section, we begin by describing the dataset used in this study and how we acquired it. Then we will move on to explaining the approach by specifying the steps needed to answer each of the four research questions.

\vspace{-0.2cm}
\subsection{Data Collection}
\label{sub:sec:dataset}
In order to conduct our qualitative and quantitative analysis, we prepared our dataset using ML projects proposed by Rzig et al. ~\cite{Rzig_Hassan_Kessentini_2022}, a subset of the dataset proposed by Gonzalez et al. ~\cite{gonzalez2020state}. Their set contained 4031 ML projects hosted on GitHub, one of the most popular platforms for hosting software repositories ~\cite{kalliamvakou2014promises}. Our set is composed of 508 open-source ML projects extracted from this set based on the following criteria. They have Python as the main programming language since it was found to be the most popular language for ML projects ~\cite{gonzalez2020state}, and use Travis CI, which is the most adopted CI/CD infrastructure in OSS projects ~\cite{zampetti2021ci} and in ML projects ~\cite{rzig2022characterizing}. 
From these projects, we extracted 15,634 commits modifying the .travis.yml file. These commits will be used for the quantitative study to answer RQ3 and RQ4.
Then, we filtered those commits based on two criteria: 
At least one Python file should be in the list of modified files and at least one of the modified source code files should have one of these keywords related to ML in their name or path:
     data, model, train, training, test, pipeline, predict, correctness, deploy, inference, preprocess. Prior works in Machine Learning ~\cite{biswas2022art,dutta2020detecting} have also utilized keyword-based searching.

The reason for this filtering is that we want to study commits that are impacting both the pipeline and ML source code. We ended up with 3169 commits, on which we applied pure random sampling with a 95\% confidence level and a 0.05 margin of error. We obtained 343 commits for manual analysis in RQ1 and RQ2.

\vspace{-0.2cm}
\subsection{Approach}
\label{sub:sec:approach}

\subsubsection{CI Evolution Analysis}\:
\label{sub:sub:sec:ci_evolution_analysis}

To understand the evolution of CI/CD pipelines, two authors manually labeled the 343 commits using  Zampetti et al.'s ~\cite{zampetti2021ci} taxonomy of restructuring actions applied in the pipeline configuration file. Both authors have extensive expertise in DevOps and Software engineering.
They focused on the second level of the taxonomy to simplify the analysis and the comparison with Zampetti et al.'s results. 
 Inter-rater agreement was measured using Cohen’s kappa~\cite{mchugh2012interrater} at 0.72, indicating substantial agreement.
 A reconciliation meeting was held afterward to resolve the differences.

\vspace{0.15cm}
\subsubsection{Co-Evolution Analysis.}\:
\label{sub:sub:sec:coevolution_analysis}

Studying the changes in the CI/CD configuration file can give us an idea of the most occurring actions performed on it. However, the intent of these changes is not obvious when they are analyzed individually.
To yield deeper insights, we analyze code changes in
both ML source files and .travis.yml file, to better understand how CI co-evolves with other ML-related components. 

\textbf{Manual co-evolution labeling.}\:
Two authors individually labeled the changes happening in the sample commits in both  ML source and .travis.yml files. Because there could be many changes occurring, the labeling was not limited to one category per commit; rather, the raters were free to add as many categories as required to assess all the changes. 
Not only did the authors analyze the code changes happening in files, but they also observed the commit message and description as well as the Pull Requests (PRs) discussion of the commit to further evaluate the adjustments made.
The labels were created by following a cooperative card-sorting procedure ~\cite{spencer2009card,miles1994qualitative}. The list of categories defined by both raters was maintained in a shared file, ensuring consistent naming without introducing substantial bias. Then, the two raters discussed and resolved conflicts in a consensual agreement meeting. We calculated the Cohen’s kappa~\cite{mchugh2012interrater} score which we found to be
0.65, showing substantial agreement. In the end, we identified 14 categories of co-changes in CI/CD pipelines and ML code. 

\textbf{Co-evolution Change Mining.}\:
To further analyze the co-evolution between CI/CD configuration changes and ML-related changes in RQ2, we employed Association Rule Mining (ARM),
which describes relationships between different phenomena. The rules consist of frequent subsets and take the form of X => Y, where X represents the antecedent and Y is the consequent.
Metrics like "Support" and "Confidence" are used to measure the quality of these rules ~\cite{alvarez2003chi}. 
Support(X => Y) indicates the frequency of appearance of co-occurrence of X and Y, while Confidence(X=>Y) measures the conditional probability that Y is present given the presence of X, indicating the reliability of the association between X and Y.

\begin{equation}
\begin{aligned}
    Supp(X \Rightarrow Y) & = P(X \cap Y)
\end{aligned}
\end{equation}
\vspace{-0.5cm}
\begin{equation}
\begin{aligned}
    Conf(X \Rightarrow Y) & = \frac{P(X \cap Y)}{P(X)}
\end{aligned}
\end{equation}
\\
To generate Association Rules, we picked the Apriori algorithm ~\cite{agrawal1994fast}, a widely used algorithm for studying co-change and co-evolution ~\cite{wang2009predicting,herzig2015empirically,marsavina2014studying}.
Given a set of transactions, the Apriori algorithm generates association rules satisfying user-specified minimum support and confidence criteria. It starts by generating large itemsets appearing in a minimum proportion of transactions using a minimum support threshold and then uses these itemsets to derive association rules. 
In RQ2, we use ARM to assess the relationship between the categories we labeled manually for co-evolution analysis. We used a minimum support value of 0.1 to focus on relatively frequent and potentially more significant associations, 
and we also used a minimum confidence value of 0.6, which was the value used in other co-evolution works~\cite{marsavina2014studying,vidacs2018co}.


\vspace{0.15cm}
\subsubsection{Change Patterns Analysis.}\:
\label{sub:sub:sec:ast_analysis}

To better understand the changes happening in CI/CD configuration files, we perform a change pattern analysis. We use Abstract Syntax Trees (AST)s, to represent the code to help mine the re-occurring changes, which we later cluster and match to generate insight.

\textbf{Abstract Syntax Tree Analysis.}\:
We are interested in studying the patterns of changes happening in the CI/CD configurations in the context of ML projects. But first, we need to parse the .travis.yml configuration file and extract the Abstract Syntax Tree (AST) ~\cite{neamtiu2005understanding}. ASTs break down code into a tree-like structure, with nodes representing different language constructs, such as functions, loops, or variables, and edges denoting the relationships between them.
To accomplish this, we used TraVanalyzer ~\cite{rzig2022characterizing}, a tool designed to parse Travis CI configuration files and generate the corresponding AST.
With it, we parse the CI/CD configuration file in all the 15,634 commits and then apply AST diffing using the state-of-the-art GumTree ~\cite{falleri2014fine} to compute fine-grained changes. 
GumTree can detect change types happening in the configuration files by comparing each node in the AST between two diffs, allowing us to pinpoint and categorize the structural changes that take place within the configuration files.

\textbf{Command Matching and Clustering of CI changes.}\:
After generating the change patterns from all the commits, we now need to find a way to group them into meaningful clusters in order to analyze the different properties and stages. However, there is an abundant use of commands and scripts within the .travis.yml file. Following Rzig et al.'s ~\cite{rzig2022characterizing} approach, we extracted the commands appearing in the file from each commit and applied a matching process where we matched AST nodes having the same command name and omitted the rest of the parameters in the commands since they are often project-specific and do not add value. 
We performed a more refined clustering based on parent node name similarity, which consists of a Travis CI keyword ~\cite{travis_keys}, and change type to generate the final list of change patterns. 
Finally, we apply a normalization procedure on the node labels to ensure the generalizability of our results.
We identified 59,948 changes in the \textit{.travis.yml} file from the dataset of commits. Each change includes the action performed, the modified command, and the parent Travis CI keywords where the command was performed.

\subsubsection{Developer Expertise Analysis}\:
\label{sub:sub:sec:developer_expertise}

Finding reliable metrics for measuring developer expertise in software development is no easy feat, as previously reported~\cite{robbes2013using,baltes2018towards}. Previous studies have employed diverse approaches, including skill vectors~\cite{dey2021representation,azad2023empirical}, ML techniques~\cite{anvik2006should,montandon2019identifying}, and, notably, Change History information ~\cite{mcdonald2000expertise,mockus2002expertise,girba2005developers}. The latter encompasses metrics such as commit frequency and the extent of modified lines of code, and has been proven effective by Anvik et al. ~\cite{anvik2007determining}.

In our study, we aim to investigate the role of developer expertise in the context of modifying CI/CD configurations for ML projects. 
We hypothesize that developers with a deeper knowledge and prolonged involvement in the project are more inclined to modify CI/CD configurations. 
In our dataset comprising 15,634 commits, we identified 1951 developers as contributors modifying pipeline configurations, with each developer uniquely identified by their email addresses. To assess expertise, we calculate the percentage of CI-modifying commits for each developer alongside the percentage of their total contributions to the projects. The objective is to explore how experience, manifested through active and substantial contributions to a project, influences the likelihood of developers being involved in CI/CD changes.
To quantify and statistically assess the strength of the observed relationships, we calculate the correlation between these two measures using Spearman's rank-order correlation ~\cite{spearman1961proof} and Kendall’s correlation ~\cite{sen1968estimates}. Spearman's correlation measures the strength and direction of the monotonic relationship between two variables. 
Kendall’s correlation quantifies the strength and direction of the ordinal association between two variables by comparing the number of concordant and discordant pairs of data points. 
We used the p-value measure to assess the statistical significance and strength of the observed correlations.
\vspace{-0.2cm}
\section{Empirical Evaluation}
\label{sec:evaluation}
\subsection{RQ1: Evolution of CI/CD pipelines}
\label{sub:sec:evolution_cicd}
\begin{figure}[!htbp]
\centering
   \includegraphics[width=0.9\linewidth]{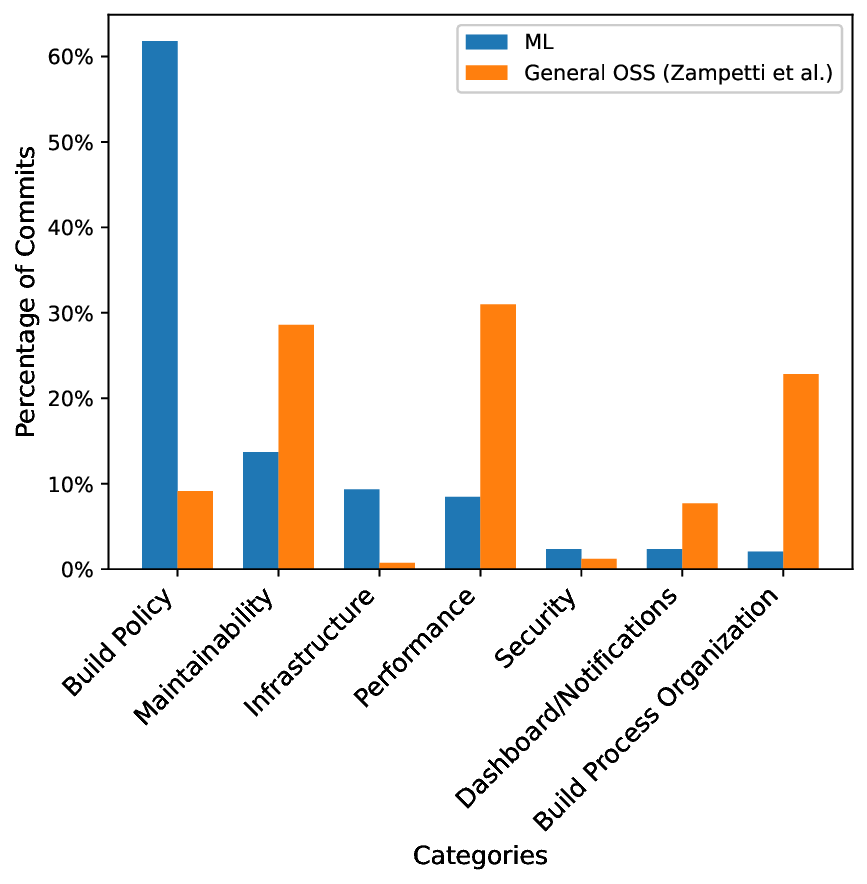}
   \vspace{-0.4cm}
    \caption{Distribution of CI/CD change categories.}
     \vspace{-0.3cm}
    \label{fig:rq1}
\end{figure}

We begin by evaluating the percentage of CI/CD pipeline changing commits in the 508 ML projects. We find a median of 4.44\%, which is slightly higher than the 2.2\% value found by Zampetti et al.'s ~\cite{zampetti2021ci} when studying general OSS projects. 
\\
To further understand how CI/CD pipelines evolve over time, 
two authors manually labeled the commits by adopting Zampetti et al.'s ~\cite{zampetti2021ci} taxonomy as described previously in ~\autoref{sub:sub:sec:ci_evolution_analysis}. This will help us in comparing the results between ML and general OSS projects, which is visualized in ~\autoref{fig:rq1}. We give a brief description of each category here, but more details are provided in Zampetti et al's ~\cite{zampetti2021ci} study.

\noindent\textbf{Build Policy} (61.8\%):
Actions in this category concern the build triggering strategy and dependencies' installation policy. We found over half of commits related to this category, with many changes found by the authors to be related to managing the dependency installation policy. This is a huge difference compared to general OSS projects as found by Zampetti et al.'s ~\cite{zampetti2021ci}. This disparity underscores the distinctive requirements and complexities associated with managing the build pipeline in ML development, highlighting the need for specialized attention in this area.

\noindent\textbf{Maintainability} (13.7\%):
Maintainability refers to the ease with which the CI/CD pipeline can be modified, repaired, and understood. Common actions include improving the readability of code snippets, renaming jobs and scripts, and simplifying the build matrix. Our analysis revealed a notable disparity in commit activity related to maintainability between ML projects and general OSS projects.
This is particularly worrisome because ML systems have a susceptibility to incurring technical debt, as they not only inherit the typical maintenance challenges associated with traditional code but also face an additional set of ML-specific issues ~\cite{sculley2015hidden}.

\noindent\textbf{Infrastructure} (9.32\%):
This category revolves around changes to the overall infrastructure supporting the build process. The identified action involves adopting Containerization to ensure a consistent environment for the build process. The two authors improved on this definition from Zampetti et al.'s ~\cite{zampetti2021ci} taxonomy to include commits where the whole \textit{.travis.yml} file was removed and replaced by a GitHub Actions configuration file. 
The raters found a very limited number of commit changes related to Containerization. ML projects often involve a diverse set of dependencies and specialized hardware configurations, making it challenging to encapsulate all aspects of the ML environment within Docker containers effectively. 

\noindent\textbf{Performance} (8.45\%):
The objective in this category is to minimize build time by removing unnecessary components from the build, adopting caching strategies, and introducing parallelization. These actions align with established practices, including recommendations from Duvall ~\cite{duvall2007continuous}.
ML models usually involve extensive computations, large datasets, and numerous dependencies, thus requiring special attention to handling these complexities when configuring CI/CD builds. However, our research revealed that ML projects did not exhibit performance improvements in CI/CD configurations to the extent anticipated. In fact, general OSS projects surpassed ML projects in terms of performance updates. This finding suggests that, despite the resource-intensive nature of ML development, general OSS projects have been more proactive in implementing optimizations to streamline their CI/CD workflows. 

\noindent\textbf{Security} (2.33\%):
This category includes actions with security implications. Examples include securing credentials/tokens in pipeline configuration and either introducing or removing \texttt{sudo} in commands. We found a low percentage of 2.33\% of changes related to security compared to other categories, which is slightly higher than that for general OSS projects.

\noindent\textbf{Dashboard/notifications} (2.33\%):
Improving the notification mechanism and enhancing build log readability in CI/CD pipelines are the main goals for this category. Here, we found a percentage of 2.33\% which is lower than the percentage for general OSS projects. Not properly configuring build output for CI/CD is considered a bad practice ~\cite{zampetti2020empirical}. Thus, ML developers need to improve on this.

\noindent\textbf{Build Process Organization} (2.04\%):
This category focuses on improving the overall organization of the CI/CD configurations through reordering the execution of build steps, restructuring install and script phases, and embracing parameterized builds. Here, we find the lowest percentage of changes which is 2.04\% compared to other categories. This is considerably lower than that of general OSS projects. 

\begin{tcolorbox}
    \textbf{RQ1 Findings:} \textit{Unlike general OSS projects, we found that over half of changes happening in CI/CD configurations are related to updating the build policy with minor changes to Performance, Maintainability, and Build Process Organization which may lead to technical debt and prolonged builds.}
\end{tcolorbox}

\subsection{RQ2: Co-evolution of CI/CD Pipelines and Machine Learning Source Code}
\label{sub:sec:test_case_effectiv}
\begin{figure}[!htbp]
    \centering
   \includegraphics[width=0.9\linewidth,height=6.5cm]{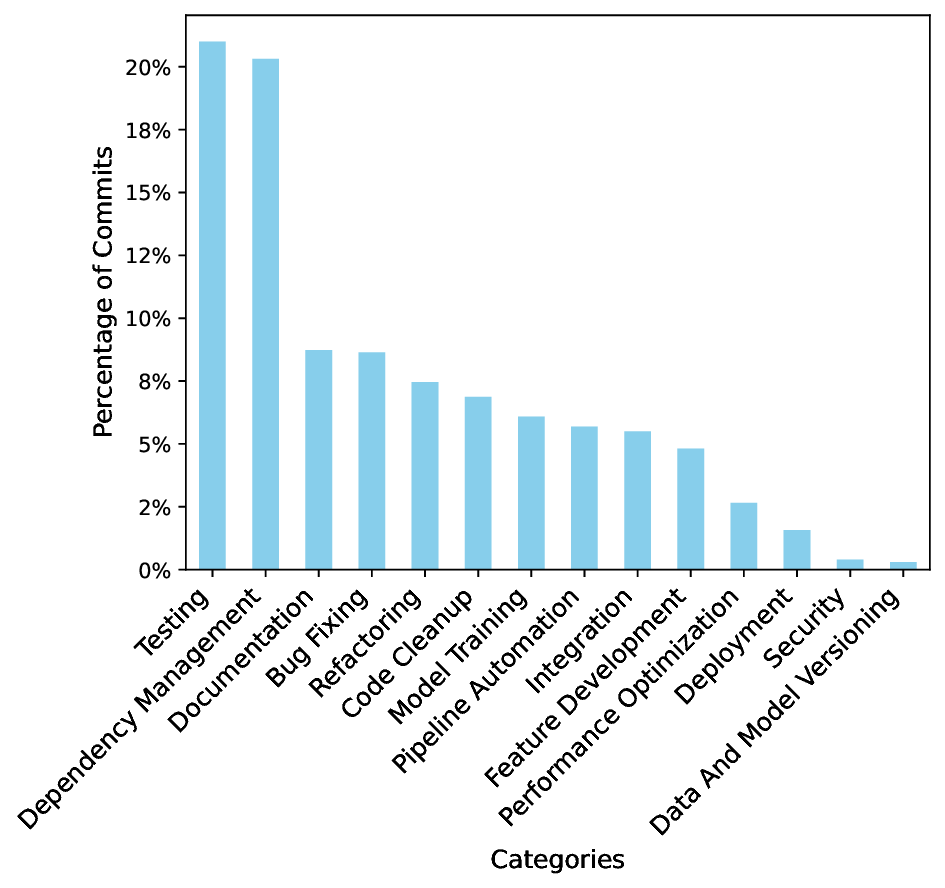}
   \vspace{-0.4cm}
    \caption{Distribution of commit categories}
    \vspace{-0.4cm}
    \label{fig:rq2_distribution}
\end{figure}

We curated a taxonomy of 14 categories to describe ML CI/CD co-changes, as explained in ~\autoref{sub:sub:sec:coevolution_analysis}. The distribution of the different categories amongst the 343 commits is shown in ~\autoref{fig:rq2_distribution}.

\vspace{-0.4cm}


We employed Association Rule Mining to assess the coupling between these different categories, which is detailed in ~\autoref{sub:sub:sec:coevolution_analysis}. We found seven association rules, which we present in ~\autoref{tab:table1}. Now we investigate the results of our analysis for each of the categories in the taxonomy.

\vspace{-0.2cm}
\begin{table}[thb]\centering
    \caption{Association rules mined for commit analysis}
    \vspace{-0.4cm}
    \label{tab:table1}
    \resizebox{0.4\textwidth}{!}{
    \large
    \begin{tabular}{*{10}{c}}
        \toprule
       Rule Antecedents & Rule Consequents &  Support &  Confidence \\
        \midrule
        Documentation & Dependency Management & 0.18 & 0.71 \\
        Pipeline Automation & Dependency Management & 0.11 & 0.67 \\
        Refactoring & Dependency Management & 0.13 & 0.61 \\
        Feature Development & Testing & 0.1 & 0.75 \\
        Integration & Testing & 0.11 & 0.71 \\
        Model Training & Testing & 0.11 & 0.66 \\
        Pipeline Automation & Testing & 0.12 & 0.72 \\
        \bottomrule
    \end{tabular}
    }
     \vspace{-0.2cm}
\end{table}

\noindent\textbf{Testing} (21.00\%):
Testing includes commits related to writing or updating tests for ML models. This is the most significant category, appearing in 21\% of all commits, meaning that approximately one in every five commits is related to testing. This supports our findings in RQ1 where we identified Build Policy, which includes updating testing policy, as the most changing category in \textit{.travis.yml} file. 
This is due to the fact that in many cases, we found that developers add the tests manually to the CI/CD configuration. An example is illustrated in ~\autoref{listing:testing}, where a new test was added to the \textit{.travis.yml} file for the ray-project/ray repository, which has over 28,000 stars. 
\begin{lstlisting}[style=yaml, caption={Adding a new test suit in \textit{.travis.yml} (ray-project/ray/d06beac).}, label={listing:testing},numbers=left,frame=lines,escapechar=\!]
script:
...
  - python test/monitor_test.py
  - python test/trial_runner_test.py
!\GreenL!+ - python test/trial_scheduler_test.py

\end{lstlisting}
This is a bad practice because each time a new test is added, the developers need also to include it in the CI/CD configuration file. The better approach is to use a testing framework with automatic discovery, like \texttt{pytest}~\cite{pytest} and \texttt{nosetests}~\cite{nosetests}, which scans for testing files and functions, making it easier to maintain and scale test suites as the project grows. 

\noindent\textbf{Dependency Management} (20.31\%):
Commits in this category involve adding or updating dependencies used in ML or CI components. The percentage of commits related to dependency management is 20.3\%, which is also about a fifth of the total commits. 
The reason for this is identified by the raters that many projects manage their dependencies within the \textit{.travis.yml} file instead of externalizing them into dedicated files, such as \textit{requirements.txt} and \textit{pyproject.toml}. This leads to frequent changes to the installation policy in \textit{.travis.yml} as we discussed in RQ1. 
For instance, we found ~\autoref{listing:dependencies} in the piskvorky/gensim repository, a well-known open-source Python library with over 14,000 stars, where dependencies are added directly in the \textit{.travis.yml} file.

\begin{lstlisting}[style=yaml, caption={Adding new dependencies directly in \textit{.travis.yml} (piskvorky/gensim/7e74d15).}, label={listing:dependencies},numbers=left,frame=lines,escapechar=\!]
install:
  ...
  - pip install scikit-learn
!\GreenL!+ - pip install tensorflow
!\GreenL!+ - pip install keras
\end{lstlisting}
Embedding dependencies in CI scripts can make it harder to maintain a consistent and reproducible environment. 

\noindent\textbf{Documentation} (8.73\%):
This category involves updating project documentation, including \textit{README} files, code comments, or API documentation. The raters found about 9\% of commits updating documentation which is moderate compared to the first two categories. Also, as shown in ~\autoref{tab:table1}, we find that if there is a documentation update within a commit, it is likely that there is also a dependency management change as well with a confidence of 0.71. This change, as observed by the two authors, is to ensure that the documentation reflects the correct dependencies and their versions. For instance in the
EducationalTestingService/skll repository, they changed the \texttt{tabulate} package to \texttt{prettytable} and had to update both \textit{.travis.yml} and the \textit{README} file.

\begin{lstlisting}[language=Python, caption={Changing the 'prettytable' dependency to 'tabulate' and updating documentation (EducationalTestingService/skll/cd5bf73).}, label={listing:documentation},numbers=left,frame=lines,escapechar=\!]
# \textit{.travis.yml}
  - conda install --yes --channel defaults --channel conda-forge python=$TRAVIS_PYTHON_VERSION numpy scipy beautifulsoup4 six scikit-learn==0.20.1 
!\RedL!-      prettytable python-coveralls ruamel.yaml
!\GreenL!+      tabulate python-coveralls ruamel.yaml
# README.rst
Requirements
...
!\RedL!-   `PrettyTable <https://pypi.org/project/PrettyTable/>`
!\GreenL!+   `tabulate <https://pypi.org/project/tabulate/>`
\end{lstlisting}

\noindent\textbf{Bug Fixing} (8.64\%):
Commits in this category address and resolve issues or bugs identified in the code. The moderate percentage might indicate that changing the \textit{.travis.yml} file and ML source code is not often related to bug fixing. Here, we observe that the bug fixes in ML files and CI/CD configurations are independent and are usually related to fixing syntax errors in \textit{.travis.yml}. 

\noindent\textbf{Refactoring} (7.46\%): 
Refactoring focuses on improving the code's structure, readability, and maintainability without altering its external behavior. ML developers might remove files, reorganize code, rename variables, or simplify complex functions to enhance the overall quality of the codebase. The category appears only in 7.46\% of commits and 
often leads to dependency updates with a confidence of 0.61. This is mostly due to removing deprecated libraries and updating Python versions, like in ~\autoref{listing:refactoring2} which was taken from the marl/openl3 repository. In this commit, developers removed Python versions 2.7 and 3.5 and added versions 3.7 and 3.8 since as part of their refactoring process, they changed their models to use Tensorflow 2 and it only supports Python versions 3.6-3.8.

\begin{lstlisting}[style=yaml, caption={Removing python versions unsupported by Tensorflow 2 (marl/openl3/d593e2d).}, label={listing:refactoring2},numbers=left,frame=lines,escapechar=\!]
python:
!\GreenL!+#- "2.7"  # byeeeee forever
!\GreenL!+#- "3.5"  # tensorflow 2 does not support
  - "3.6"
!\GreenL!+ - "3.7"
!\GreenL!+ - "3.8"
\end{lstlisting}

\noindent\textbf{Code Cleanup} (6.87\%):
Commits in this category are related to removing dead code, unused variables, or deprecated functions to help maintain a clean and efficient codebase. We find a percentage of 6.87\% commits which is moderate compared to other categories. We show a code snippet reflecting this category from the chainer/chainercv repository, where they removed the disk attribute, a decorator that marks tests consuming a lot of disk space, from the \textit{.travis.yml} file as well as source files as shown in ~\autoref{listing:codecleanup}.

\begin{lstlisting}[language=Python, caption={Removing the disk test decorator from test files and \textit{.travis.yml} (chainer/chainercv/3eff205).}, label={listing:codecleanup},numbers=left,frame=lines,escapechar=\%, literate={@}{{\textcolor{black}{@}}}1]
# \textit{.travis.yml}
%\RedL%- MPLBACKEND="agg" nosetests -a '!gpu,!slow,!disk' tests;
%\GreenL%+ MPLBACKEND="agg" nosetests -a '!gpu,!slow' tests;
# tests/datasets_tests/ade20k_tests/test_ade20k.py
%\RedL%-    @attr.disk
     def test_ade20k_dataset(self):
\end{lstlisting}

\noindent\textbf{Model Training} (6.08\%):
Model Training commits are related to training or fine-tuning ML models. Changes to hyperparameters, datasets, or training algorithms fall under this category. Model training is essential for optimizing the performance of ML applications. However, with a percentage of approximately 6\%, we realize that changing model behavior is not always correlated with updating CI/CD configurations. 
We also found that model training often requires adding new tests and/or updating older ones with a confidence of 0.66. An example of that is shown in ~\autoref{listing:modeltraining} taken from the OpenNMT/OpenNMT-py. Here, the developers decided to use training steps instead of epochs when training the models. Training steps provide more fine-grained control over the training process. A change in the testing command was performed in \textit{.travis.yml} as well.

\begin{lstlisting}[language=Python, caption={Using training steps instead of epochs for model training (OpenNMT/OpenNMT-py/4d17982).}, label={listing:modeltraining},numbers=left,frame=lines,escapechar=\!]
  - python train.py -model_type img -data /tmp/im2text/q -rnn_size 2 -batch_size 10 -word_vec_size 5
!\RedL!-       -report_every 5 -rnn_size 10 -epochs 1 
!\GreenL!+       -report_every 5 -rnn_size 10 -train_steps 10
\end{lstlisting}

\noindent\textbf{Pipeline Automation} (5.69\%):
These commits introduce or enhance automation scripts or workflows within the CI/CD pipeline. Automated processes improve efficiency and reliability, enabling seamless integration, testing, and deployment of machine learning models. The percentage of 5.69\% commits suggests minimal efforts for using automation scripts within these commits. 
Furthermore, We noted that adding automation scripts often coincides with updates in testing components and changing dependencies as shown in ~\autoref{tab:table1}. 

\noindent\textbf{Integration} (5.50\%):
Integration commits signify connections with other systems or services, aligning with a broader software ecosystem. Integration can include databases, web applications, or notification systems. We found a percentage of 5.5\% commits which is moderately low compared to other categories. 
We also found that these changes are frequently followed by updates in testing components with a confidence of 0.71 in ~\autoref{tab:table1}. This indicates a strong correlation between integration efforts and ensuring comprehensive testing, possibly to validate the integrated systems.
The raters observed that in most cases, the service is actually Docker and is used to run integration tests as shown in ~\autoref{listing:integration}, which was taken from the X-DataInitiative/tick repository.

\begin{lstlisting}[style=yaml, caption={Integrating Docker to build and run tests (X-DataInitiative/tick/01a4966).}, label={listing:integration},numbers=left,frame=lines,escapechar=\!]
matrix:
  include:
...
!\GreenL!+   services: docker
!\GreenL!+   env:
!\GreenL!+     - DOCKER_IMAGE=xdatainitiative/tick_ubuntu:1.3
...
script:
...
!\GreenL!+ - if [[ "$TRAVIS_OS_NAME" == "linux" ]]; then docker 
!\GreenL!+     run -v `pwd`:/io "$DOCKER_IMAGE" /io/tools/travis/
!\GreenL!+     docker_run.sh; fi
\end{lstlisting}

\noindent\textbf{Feature Development} (4.81\%):
Feature Development involves the addition of new ML features to the codebase. 
This could include implementing novel algorithms, data processing techniques, or any other functionality that enhances the capabilities of the machine learning models. The low frequency suggests that new features do not require updates in the CI/CD configurations.
However, those changes are highly associated with testing updates with 0.75 confidence as shown in ~\autoref{tab:table1}. 

\noindent\textbf{Performance Optimization} (2.65\%):
Commits aiming to optimize the performance of machine learning models or CI/CD processes fall into this category. The techniques used are mentioned in RQ1. 

\begin{lstlisting}[language=Python, caption={Using the joblib~\cite{joblib} library for parallel computing (tslearn-team/tslearn/d3062d3).}, label={listing:performance},numbers=left,frame=lines,escapechar=\!]
# \textit{.travis.yml}
  - conda create -q -n test-environment python=$TRAVIS_PYTHON_VERSION Cython numpy>=1.14 scipy tensorflow keras
!\RedL!-       scikit-learn numba nose
!\GreenL!+       scikit-learn numba joblib>=0.12
# tslearn/clustering.py
def silhouette_score(X, labels, metric=None, sample_size=
!\RedL!-  None, metric_params=None, random_state=None,
!\GreenL!+  None, metric_params=None, n_jobs=None, random_state=
!\GreenL!+  None,**kwds):
\end{lstlisting}

The raters found a few cases involving using parallelization through \texttt{joblib}~\cite{joblib}, a well-known Python library that provides tools for parallel computing and efficient caching.
~\autoref{listing:performance} is a code snippet taken from the tslearn-team/tslearn repository where the developer added used the joblib package to perform parallel computations within their models. They added the dependency to the installation command in \textit{.travis.yml} since it is now needed to build and test the models.

\noindent\textbf{Deployment} (1.57\%):
Deployment commits involve activities related to deploying machine learning models to production or updating deployment-related code. 
Surprisingly, we found a small number of commits that apply deployment changes when modifying the \textit{.travis.yml} file. 
ML practitioners seem to prefer more manual and controlled deployment processes, reflecting a risk-averse attitude toward automated deployment. The limited emphasis on CD could stem from the unique challenges posed by ML models, emphasizing the need for precision and careful consideration in better-tailored deployment processes for ML projects.

\noindent\textbf{Security} (0.39\%):
This category addresses security vulnerabilities or improves the security aspects of machine learning models, data handling, or CI/CD processes. With only four commits related to security, it is obvious that security is not a major concern when updating both ML source code and pipeline configurations. This aligns with our finding from earlier in RQ1, where we also identified only eight commits related to security in \textit{.travis.yml}.

\noindent\textbf{Model And Data Versioning} (0.29\%):
These commits are related to data and model versioning and management, ensuring consistency and reproducibility in ML experiments. This category has the lowest frequency with only 3 commits. 
Although versioning is established as a good practice, ML versioning is still a young practice as observed by Lewis et al. ~\cite{lewis2021software}, due to the lack of effective tools tailored for the complexity of ML models. Traditional code versioning tools like Github are unsuitable for data versioning due to large sizes and specialized tools like Data Versioning Control (DVC) are still not widely adopted in ML projects, as found by Barrak et al. ~\cite{barrak2021co}. Two of the commits we found just use Github to specify datasets with different versions whilst one commit uploads the model to Amazon Web Services (AWS)~\cite{aws} storage with their corresponding versions as a workaround for the large file sizes. The commit in ~\autoref{listing:versioning}, from sorgerlab/indra,  updates the version of the folder containing the models in the cloud storage and renames to reflect the version update from 1.2 to 1.3.

\begin{lstlisting}[language=Python, caption={Versioning data in AWS storage (sorgerlab/indra/6bba5a2).}, label={listing:versioning},numbers=left,frame=lines,escapechar=\!]
!\RedL!-  - mkdir -p $HOME/.indra/bio_ontology/1.2
!\GreenL!+  - mkdir -p $HOME/.indra/bio_ontology/1.3
!\RedL!-  - aws s3 cp s3://bigmech/travis/mock_ontology.pkl 
!\RedL!-    $HOME/.indra/bio_ontology/1.2/
!\GreenL!+  - aws s3 cp s3://bigmech/travis/bio_ontology/1.3/
!\GreenL!+    mock_ontology.pkl $HOME/.indra/bio_ontology/1.3/
     bio_ontology.pkl --no-sign-request
\end{lstlisting}

\begin{tcolorbox}
    \textbf{RQ2 Findings:} \textit{We devised a taxonomy of 14 co-changes and identified Testing and Dependency Management as the most prominent categories. We found two bad practices in those two categories which are direct inclusion of dependencies and a lack of usage of standardized testing frameworks.}
\end{tcolorbox}

\subsection{RQ3: Change Patterns in CI/CD pipelines}
\label{sub:sec:change_patterns}

\begin{figure}[!htbp]
    \centering
   \includegraphics[width=\linewidth]{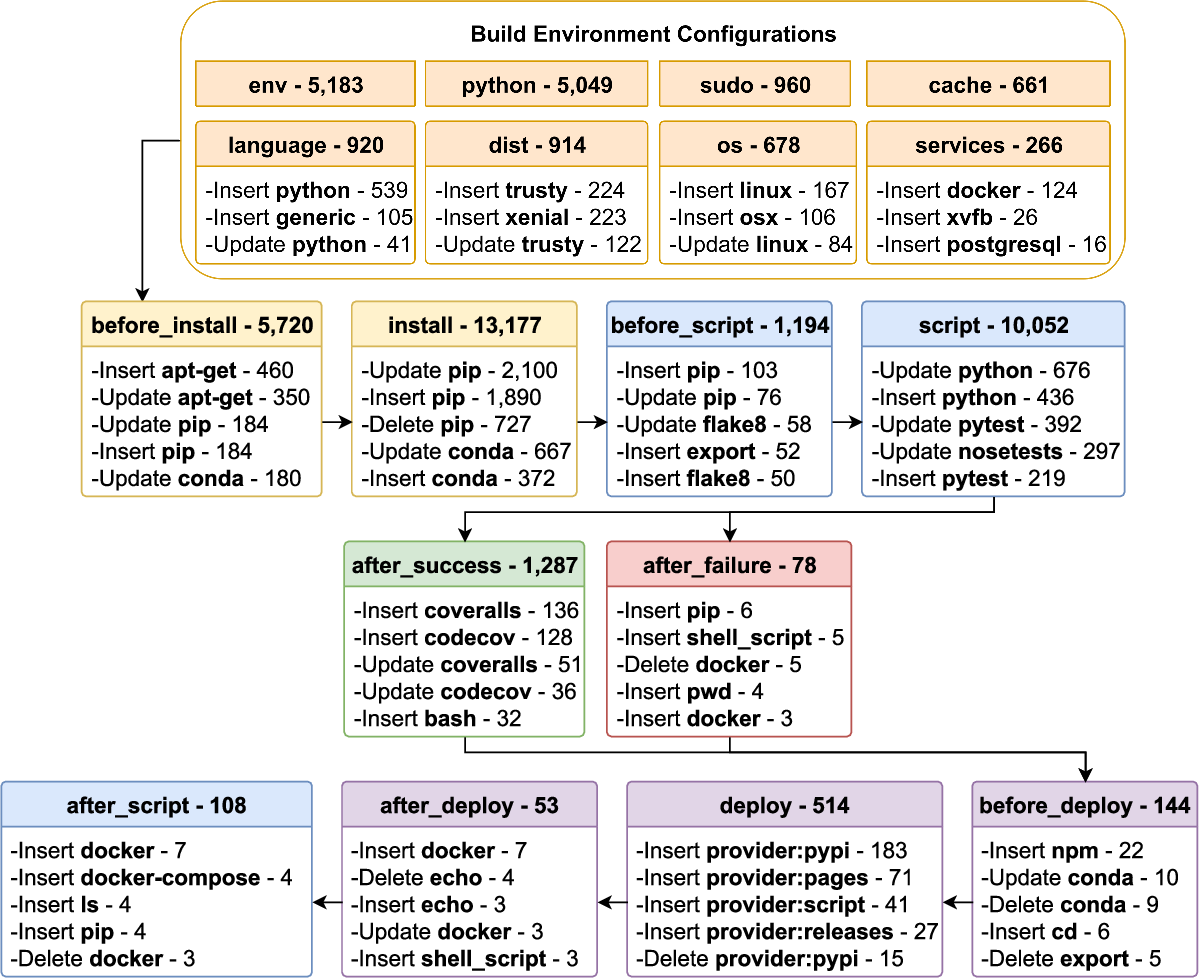}
   \vspace{-0.5cm}
    \caption{Change Patterns in \textit{.travis.yml} configurations lifecycle.}
    \label{fig:rq3_lifecycle}
    \vspace{-0.2cm}
\end{figure}

To gain a deeper understanding of the evolution of CI/CD configurations, we want to identify the change patterns occurring within build environment settings and the different job phases.
To achieve that, we perform AST analysis on \textit{.travis.yml} configuration files and then apply clustering and matching to obtain a list of change patterns, as explained in ~\autoref{sub:sub:sec:ast_analysis}.
The results of this procedure are shown in ~\autoref{fig:rq3_lifecycle} where we present each phase and its frequency as well as the top five change patterns within each one.

\noindent\textbf{Build Environment Configurations}:
As shown in ~\autoref{fig:rq3_lifecycle}, we identified the top eight build environment configurations and the frequency of their changes. For some of them, we also specified the top three change patterns. 
Setting environment variables via the \texttt{env} key is the most changed configuration with 5,183 occurrences. By adjusting environment variables, developers can easily specify and control the test environment. 
Furthermore, it provides flexibility for testing since ML projects often involve diverse frameworks, data sources, or experimental conditions.
The next most changed setting is related to updating Python versions. 
Developers may want to ensure their projects are tested across various Python environments to guarantee broad compatibility. 

We also found 960 updates to the \texttt{sudo} key which is moderately low compared to other patterns. This aligns with our results in RQ1 and RQ2 where we found that security is not a big concern for ML developers.
However, we found that the \texttt{sudo} key is actually deprecated according to Travis CI documentation ~\cite{travis_keys}. It is a bad practice in software engineering to use deprecated code.
Additionally, we identified 661 changes related to caching, which is also moderately low compared to other keys. As explained in RQ1 and RQ2, performance-related changes are minimal in \textit{.travis.yml} file within ML projects. 

The \texttt{language} key has been updated 920 times in our dataset, and most changes are related to adding or updating the Python language or using a generic language. The generic language means that the build environment should not assume a specific programming language and should provide a generic environment where developers can specify their own build commands. However, since the projects we are working with are all written in Python, we find this behavior unsettling. Using "python" as the language key provides a more standardized and optimized environment.
The \texttt{os} and \texttt{dist} keys were modified moderately with \texttt{linux} and \texttt{osx} being the most changed operating systems.

As for distribution, most changes are related to adding either the \texttt{trusty} distribution or the \texttt{xenial one}. Both represent a version of the Ubuntu operating system with specific features and package versions. 

Finally, we found minimal changes related to integration with services, a detail we also observed in RQ2. Similarly to our findings in RQ2, most changes are related to adding \texttt{docker}. However, as observed in RQ1, there are minimal changes related to Containerization due to the difficulty of encapsulating the complex aspects of the ML environment within Docker containers.
Other added services include \texttt{xvfb}, a service that provides a virtual display server for running graphical applications, and PostgreSQL, a pre-configured environment that provides a running instance of the PostgreSQL database server for testing. 
The usage of the \texttt{xvfb} service in CI/CD configuration within ML projects appears somewhat unconventional due to the non-graphical nature of many ML tasks.

\noindent\textbf{Job Phases}:
We observe that the \texttt{install} phase has undergone the most changes with a frequency of 13,177. The top five change patterns we found are related to 
\texttt{Pip} and \texttt{Conda} which are considered the most popular package managers ~\cite{shaffer2021empirical}. 
Additionally, the \texttt{before\_install} phase is the third most changed phase as well which is often used for pre-setup tasks before the main installation phase. This supports the results we found in RQ1 and RQ2, where we identified updating dependencies as the second most frequent category of change happening in CI/CD changing commits as well as updating installation build policy as the most occurring action in \textit{.travis.yml} file.

The \texttt{script} phase is the second most frequently changed. This phase typically includes commands for running tests, as shown in the change patterns in ~\autoref{fig:rq3_lifecycle}. The high number of changes indicates a significant amount of activity related to test scripts, which we observed in RQ2. 
\texttt{Pytest}~\cite{pytest} and \texttt{nosetests}~\cite{nosetests} are popular testing frameworks for Python but surprisingly, the python command is changed more frequently, potentially indicating a tendency to run tests independently using the python command rather than utilizing standard testing frameworks, a bad practice that we also noted in RQ2.
As for the \texttt{"before\_script"}, it is also frequently changed. Surprisingly, we found that the pip command is changed constantly here. In Travis CI configurations, it's more common and considered best practice to use the install phase for installing dependencies. We also found the flake8 command, often used to perform linting and static code analysis, and the export command which sets environment variables. Updating these commands aligns with the idea of performing necessary setup and checks before the main build or testing phases commence which is the intent of the \texttt{before\_script} phase. 

The discrepancy between the frequency of changes in the \texttt{after\allowbreak\_success} and \texttt{after\_failure} phases may be attributed to the distinct nature of these phases within the CI/CD pipeline. \texttt{after\allowbreak\_success} typically includes actions performed when the build and tests have passed successfully, indicating a stable state, whereas \texttt{after\_failure} is executed in the event of test failures or build errors. 
Changes in \texttt{after\allowbreak\_success} are mostly related to adding the \texttt{coveralls} and \texttt{codecov} commands, which report code coverage metrics to external services after a successful build.
The \texttt{after\_failure} less frequent usage can indicate that ML developers often resort to manual intervention or debugging outside the CI/CD pipeline when errors occur, thereby reducing the need for frequent adjustments. 

Furthermore, we note that deployment-related phases, which are \texttt{before\allowbreak\_deploy}, \texttt{deploy} and \texttt{after\_deploy} have low frequencies compared to other phases which suggests a cautious approach to automated deployment in the ML community. We found similar results in RQ2 as well with Deployment being one of the least frequent categories of changes. The most commonly added provider is PyPI, followed by a few others such as GitHub Pages and releases, as well as custom script-based deployment strategies. 

\begin{tcolorbox}
    \textbf{RQ3 Findings:} \textit{We generated a comprehensive list of change patterns. analyzing those supports our findings in RQ1 and RQ2. We observe two more bad practices when updating CI/CD configurations which are the usage of deprecated Travis CI settings and the reliance on a generic build language.}
\end{tcolorbox}

\subsection{RQ4: Developer Expertise for CI/CD Configuration Changes}
\label{sub:sec:dev_expertise}
Developer expertise has been a well-explored area in the context of recommendation systems, with substantial research highlighting its significance ~\cite{mcdonald2000expertise,mockus2002expertise,minto2007recommending}.
However, it's worth noting that, to the best of our knowledge, there is a noticeable gap in research concerning developer expertise in the domain of CI/CD pipelines for ML projects.
As explained in ~\autoref{sub:sec:dev_expertise}, we used change history, and specifically the frequency of commits, as a metric to evaluate the experience level needed to conduct changes in CI/CD configurations.
\begin{figure}[!htbp]
\centering
   \includegraphics[width=0.9\linewidth]{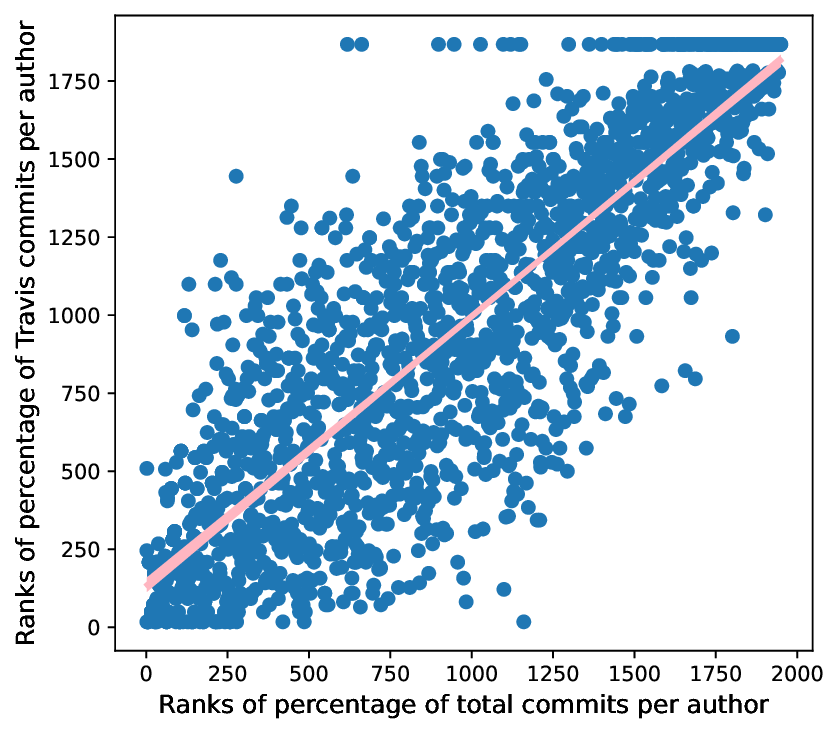}
   \vspace{-0.4cm}
    \caption{Ranks of percentage of commits per author (Spearman's)}
     \vspace{-0.4cm}
    \label{fig:rq4}
\end{figure}
We calculate the correlation between the percentage of CI-modifying commits for each developer and the percentage of their overall contributions to the projects using Spearman's correlation ~\cite{spearman1961proof} and Kendall’s correlation ~\cite{sen1968estimates}.  
We observed a Spearman's correlation value of 0.86. The associated p-value was extremely small (p < 0.001), indicating strong evidence against the null hypothesis of no correlation. 
This significant correlation is visually represented in ~\autoref{fig:rq4} using a scatter plot of the ranked values. We also found Kendall's correlation value of 0.68 with an extremely low p-value (p < 0.001).
These results show that the developers responsible for modifying the CI/CD configuration file are likely to be the ones contributing the most to the ML projects, and thus should have the most expertise in those projects.


\begin{tcolorbox}
    \textbf{RQ4 Findings:} \textit{
    Developers with deeper knowledge and prolonged involvement in the project are more inclined to modify CI/CD configurations.}
\end{tcolorbox}

\vspace{-0.4cm}
\section{Threats to Validity}
\label{sec:discussion}
\noindent\textbf{Construct validity.} 
The manual analysis of RQ1 and RQ2 might be insufficient to evaluate the exact changes happening in the commit. Since there is no existing taxonomy to use as a reference, the two authors independently reviewed and categorized the changes in the sample commits. The categorization process took into account various sources, including code diffs, commit messages, commit descriptions, and pull request discussions. Then, the authors held a meeting to discuss the conflicts and reach a consensus. 
Additionally, for assessing developer expertise in RQ4, where no exact mean of measurement exists, we leveraged change history information, a reliable metric well-documented in the literature ~\cite{mcdonald2000expertise,mockus2002expertise,girba2005developers,anvik2007determining}.

\noindent\textbf{Internal validity.} 
We acknowledge the potential for selection bias in our sample dataset, which may not fully represent all changes in CI/CD commits. We reduce this threat by using random sampling with a 95\% confidence level and a 0.05 margin of error when creating our dataset. 
Furthermore, The observed coupling between commit categories can be due to noise. To mitigate this threat, minimum support and minimum confidence thresholds were applied. 

\noindent\textbf{External validity.} 
Our study focused on open-source ML projects using Python as their primary programming language, which is the most popular language for ML projects ~\cite{gonzalez2020state}. We acknowledge the limitation of generalizing our findings to closed-source projects and those developed in different programming languages. To address this limitation, our dataset was diverse, encompassing projects of varying sizes, ages, and commit frequencies. 
Furthermore, we only studied projects that have Travis CI as their CI/CD infrastructure, which is considered the most used CI/CD tool for open-source ML projects ~\cite{rzig2022characterizing}.

\section{Related Works}

\label{sec:related}
\subsection{CI/CD Bad Practices and Barriers}
The challenges of adopting CI/CD pipelines have been highlighted by many authors. Duvall et al.~\cite{duvall2007continuous} first identified several common barriers related to using CI/CD pipelines such as maintenance, managing dependencies, and handling different environments. He then curated a catalog of 50 patterns and anti-patterns regarding several phases in the CI process~\cite{duvall2011continuous}. Zampetti et al.~\cite{zampetti2020empirical} also defined a catalog of 79 bad smells encountered by developers, leveraging interviews with experts and analyzing Stack Overflow posts. Hilton et al.~\cite{hilton2017trade} studied the challenges faced by developers when moving to CI, which involve multiple aspects such as quality assurance, security, and flexibility. Similarly, Olsson et al.~\cite{olsson2012climbing} examined the barriers of migration towards CD.
Our work is focused more on analyzing the co-changes occurring in CI/CD and ML code.

\subsection{CI/CD in Machine Learning Projects}

There is limited literature studying CI/CD usage within ML projects. Some works~\cite{karlavs2020building,renggli2019continuous} found the traditional testing practices in existing CI services to be insufficient when it comes to ML applications and proposed new CI systems more tailored to the specifications of ML testing specifications. Rzig et al.~\cite{rzig2022characterizing} was one of the first researchers to empirically study and characterize CI adoption rate, performed tasks, and build failures in ML projects compared to general OSS projects. However, his work mainly focused on analyzing CI adoption without delving deeper into the changes occurring in CI/CD configuration and the developer expertise needed to perform those modifications.

\vspace{-0.2cm}
\subsection{Software Evolution}
Many papers have studied the evolution of software artifacts. McIntosh et al.~\cite{mcintosh2011empirical} empirically studied the evolution of build systems in open-source projects and found that build files have a high churn rate and are tightly coupled with source code and test files, which means that they need constant maintenance as the source files and test files changes. They also studied the developer efforts and found that 79\% of source code developers also change build files. 
Jiang et al.~\cite{jiang2015co} explored the co-evolution of Infrastructure-as-Code (IaC) files and found IaC files to be tightly coupled with other software artifacts. 
Barrak et al.~\cite{barrak2021co} focused on the co-evolution of Data Versioning Control (DVC) files and ML source files, and found a tight coupling between DVC and software artifacts and a non-constant complexity trend for DVC files in 78\% projects.
Zampetti et al.~\cite{zampetti2021ci} studied the evolution of CI/CD pipelines by evaluating the restructuring actions occurring in the CI/CD changes. The 34 restructuring actions are organized in a three-level taxonomy and helped in extracting 16 metrics describing how pipelines evolve over time. 
Unlike Zampetti et al.~\cite{zampetti2021ci}, we want to understand the co-evolution between changes happening in the CI/CD pipeline configuration and the ones happening in ML source code. We also analyzed the developer expertise needed to perform those changes.
\section{Implications}
\label{sec:implications}

\noindent\textbf{For ML Developers:}
The study's findings underscore the critical importance of managing dependencies and testing procedures in ML projects, as these areas experience frequent changes and often need adjustments in build policies. ML developers need to pay extra care in these areas and try to avoid bad practices like managing dependencies directly in CI/CD configuration and not using standardized testing frameworks. These practices can lead to CI/CD maintenance overhead and bugs. Furthermore, as ML projects tend to use large datasets and complex computations, ML developers need to utilize caching mechanisms and job parallelization to enhance the performance of the CI build.

\noindent\textbf{For ML Tool Builders:}
The study highlights a significant opportunity for tool developers to streamline the CI/CD process for ML developers. The limited adoption of continuous deployment among ML developers when updating CI/CD presents a valuable opportunity for tool builders to develop and provide solutions tailored to the specific needs and challenges of the machine learning development workflow. 
Furthermore, building upon our identified change patterns in RQ3, there is an opportunity for CI tools to become more tailored to ML projects. This could involve enhanced documentation features to assist in the creation of CI configuration files tailored to the specific needs of ML development. Additionally, incorporating prompts with commonly used commands could facilitate the onboarding of less experienced developers, thus mitigating the perception that only experts can effectively modify CI/CD files, as observed in our earlier findings.
Our dataset of change patterns from RQ3 can also be used for improving static analysis tools.

 \noindent\textbf{For Researchers:}
Our findings reveal a prevalence of bad practices among ML developers in CI/CD processes, presenting an opportunity for researchers to delve into this domain. Researchers can leverage the existing list of change patterns to conduct in-depth investigations into code smells and bad practices within ML projects. This approach allows for the development of tailored guidelines and best practices aimed at improving the overall quality and efficiency of CI/CD workflows within the machine learning development ecosystem.

\section{Conclusion}
\label{sec:conclusion}
 In this paper, we presented the first empirical analysis of how CI/CD configuration changes and co-evolves with ML code during the life cycle of ML projects. Moreover, we performed CI/CD change pattern analysis and evaluated the expertise of ML developers who manage CI/CD configurations. 
 Our analysis found that over half of commits include updates to the build policy and minor changes related to performance and maintainability compared to general open-source projects.
 We also revealed several bad practices performed by ML developers which include managing dependencies directly in CI/CD files, using deprecated code, and not utilizing standardized testing frameworks.
 Moreover, the pattern analysis identified common integration and delivery features widely used in different CI/CD execution phases. 
 At the same time, our developer expertise for CI/CD maintenance identified that the pipeline is mostly managed by experienced developers, which indicates limited knowledge of CI/CD among the ML development community. 
 We hope that our findings on CI/CD change analysis on ML projects will allow future researchers to develop techniques for automatic incorporation and synchronization of the CI/CD pipeline for ML projects.
\begin{acks}
  The UofM-Dearborn authors are supported in part by UofM-Dearborn Research Support and NSF Award NSF-2152819.
\end{acks}

\bibliography{biblio}

\end{document}